\documentclass[12pt,notitlepage, nofootinbib,
longbibliography]{revtex4-1}
\usepackage{graphicx}
\usepackage{color}
\usepackage{amsfonts}
 
\usepackage{graphicx}
\usepackage{color}
\usepackage{amsfonts}

\begin{document}

\title[Bell states via repeated measurements]
{Generation and stabilization of Bell states via repeated projective measurements on a driven ancilla qubit
}

\author{L.~Magazz\`u$^{1}$, J.~D.~Jaramillo$^{1}$, P.~Talkner$^{1}$ and P.~H\"anggi$^{1,2}$}

\address{$^{1}$Institute of Physics, University of Augsburg, D-86135 Augsburg, Germany}
\address{$^{2}$Nanosystems Initiative Munich, Schellingstra{\ss}e 4, D-80799 M\"unchen, Germany}
%\ead{}

\date{\today}

\begin{abstract}
A protocol is proposed to generate Bell states in two non-directly interacting qubits by means of repeated measurements of the state of a central ancilla connected to both qubits. 
An optimal measurement rate is found that minimizes the time to stably encode a Bell state in the target qubits, being of advantage in order to reduce detrimental effects from possible interactions with the environment.
The quality of the entanglement is assessed in terms of the concurrence and the distance between the qubits state and the target Bell state is quantified by the fidelity. 
\end{abstract}


% Some applications: cryptography, dense coding, teleportation, quant repeaters, entanglement swapping, quant memories, metrology, etc.

\maketitle

\section{Introduction}
 
\begin{figure}[htbp]
   \centering
   \includegraphics[width=2.8in]{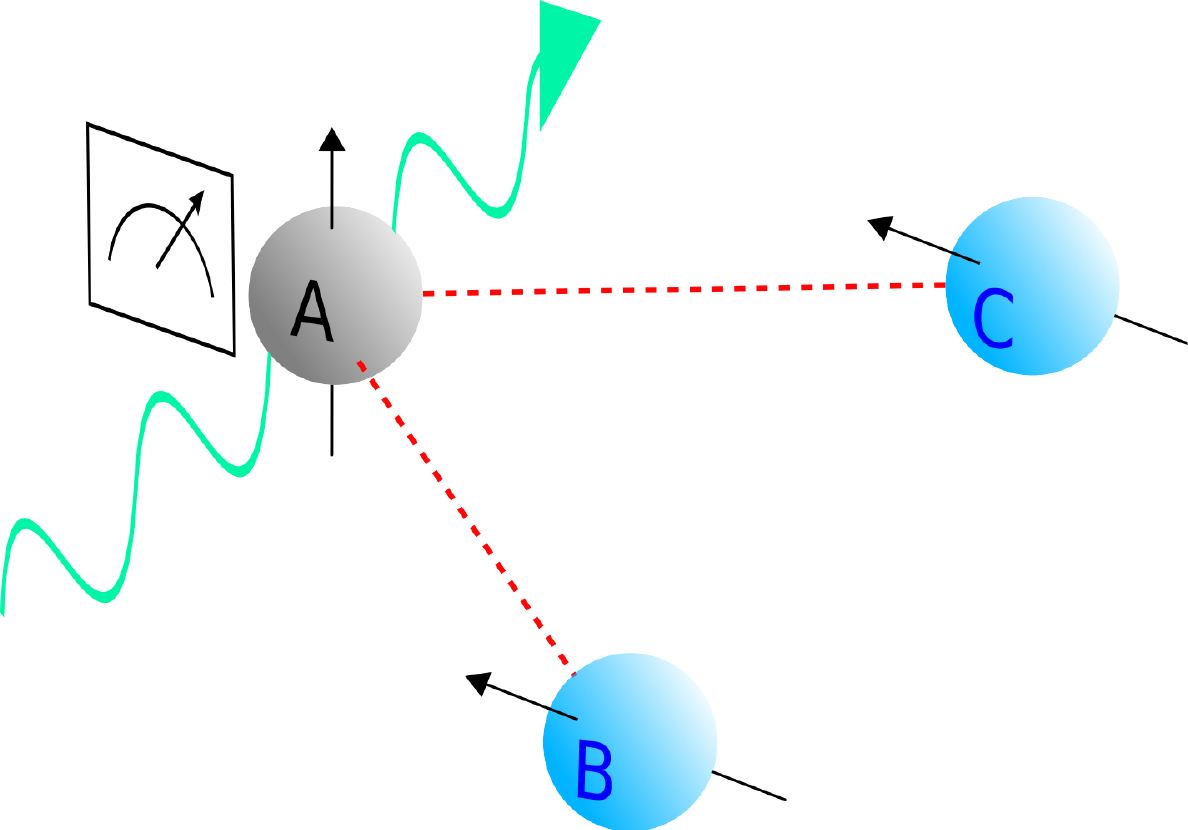} 
   \caption{Two mutually non-interacting qubits ($B$ and 
   $C$) are coupled to a shared ancilla $A$ whose  state ($\sigma_z$) is projectively monitored. During the protocol, a control field acts on $A$ triggered by a specific readout. We assume the qubits to be isolated from the environment. }
   \label{fig:setup}
\end{figure}

Preparing entangled states is a basic requirement for many quantum technologies \cite{Schleich2001,Vuckovic09},
 notably for quantum information \cite{Horodecki09} and quantum metrology \cite{Maccone11,Apellaniz14}.  Being entanglement an exquisite  non-classical feature, its quantification  is of fundamental interest~\cite{Wootters97,Wootters98, Schleich2006}. 
Several protocols to generate entangled states have been developed to date, including  control of quantum dynamics \cite{Martinis06,Hanggi09a,
Hanggi09b,Hanggi14} and  engineered dissipation \cite{Huelga02,Cirac09,
Kohler10,Li12,Mirrahimi13,
Siddiqi16,Giedke16,Sorensen16,
Li17,Li2017}.
An intriguing route towards this goal is to exploit the quantum backaction of measurements performed on a part or on the whole system.
% to produce entanglement between its constituents.
In this context, different schemes have been proposed~\cite{Beenakker2004,Loss2005, Korotkov03,Buttiker06, Kolli2006,Ralph08,Jordan08}  and implemented~\cite{Hanson12,DiCarlo14,Siddiqi2014} 
which rely on the use of a parity meter on the collective state of two qubits. The introduction of a feedback control based on the readout of a continuous weak measurement of parity provides further means to entangle bipartite systems~\cite{Siddiqi2016,Martin2017, DiCarlo13,Romito14}. 
A parity meter of the state of two qubits, $\alpha$ and $\beta$, discriminates if they  are in an even or odd parity collective state, associated to the two eigenvalues $1$ and $-1$ of the parity operator $\sigma_z^\alpha\otimes\sigma_z^\beta$, respectively. Consider the one-qubit state $|+\rangle=(|\uparrow\rangle+|\downarrow\rangle)/\sqrt{2}$ expressed in the eigenbasis of $\sigma_z$. 
A parity measurement on the two-qubit system prepared in the separable joint  state $|+_\alpha\rangle|+_\beta\rangle$    
projects the system onto one of the Bell states $|\Phi^+\rangle=(|\uparrow\uparrow\rangle+|\downarrow\downarrow\rangle)/\sqrt{2}$ and $|\Psi^+\rangle=(|\downarrow\uparrow\rangle+|\uparrow\downarrow\rangle)/\sqrt{2}$, corresponding to   even and odd parity outcome, respectively. 
 An advantage of the class of  schemes based on parity measurements is that they do not require direct interaction between the qubits, a feature that makes them suitable for linear optics setups, e.g., the scheme for quantum computing discussed in Ref~\cite{KLM2001}.\\ 
\indent In practical implementations, the parity measurements may involve the coupling to an ancillary qubit  upon which measurements are performed, and the use of multi-qubit gates to prepare the ancilla qubit as a parity meter~\cite{Hanson12,DiCarlo14}. 
On the other hand, the action  on the ancilla to drive the system state, allows for keeping the target qubits more isolated from the environment (which in general includes the measurement apparatus). In such a situation, the ancillary system can thus be considered as a so-called quantum actuator (see Ref.~\cite{Kempf2016} and references therein) accomplishing the indirect control of the system state.\\ 
\indent In the present work we exploit the idea of a shared ancilla driven by the measurement backaction~\cite{Nakazato2004,Wu2004, Compagno2004} to circumvent the use of collective unitary gates to generate Bell states in a bipartite target system.
Specifically, we encode and stabilize the Bell states in a couple of mutually noninteracting qubits ($B$ and $C$) by repeated projective measurements of the state of a shared ancilla qubit ($A$) which may also driven by local control fields (see Fig.~\ref{fig:setup}). The sequence of measurements is performed starting with the full system in the factorized state with the three qubits in the same spin state. A readily implementable form of feedback, i.e., a ramp of the control field on $A$ triggered by a specific outcome of the measurement, ensures that in the ideal case of perfect isolation from the  environment the protocol yields a Bell state with probability 1. Moreover, the sequence of outcomes of the measurements on $A$, unambiguously identifies the specific Bell state in which the target qubits $B$ and $C$  are left asymptotically. One can then switch among the four Bell states encoded in $BC$ by addressing locally either $B$ or $C$ with a single qubit  operation~\cite{Preskill2004}, an action that does not require proximity or interaction between the entangled qubits. The feasibility of our scheme benefits from the progress in rapid high-fidelity, single-shot readout in circuit QED, superconducting qubits, and spins in solid state systems~ \cite{Esteve09,Dzurak10, Jelezko10,Hanson11, Morello13,Imamo14, Bylander16,Wallraff17,Petta2017}.\\  
\indent By analysing the results of the protocol with different inter-measurement times, we are able to identify the optimal rate of measurements to generate stable Bell states in a minimal amount of time. This is crucial for successfully  producing a Bell state before the detrimental effects of environmental noise spoil the protocol~\cite{Wong-Campos2017}. We then use this optimal rate to examine how stable Bell states in the target qubits can be produced in a short time window, comprising $\sim 20$ measurements on ancilla $A$, before the latter is disconnected leaving the target qubits isolated. We explicitly depict sample time evolutions of the density matrix, corresponding to different realizations of the measurement sequences.
 
 \section{Setup}
 
The model considered in our protocol consists of an open one-dimensional chain comprised of three $1/2$-spins  $A$, $B$, and $C$ (the qubits, see Fig.~\ref{fig:setup}). The central spin $A$ plays the role of an ancilla and is connected to the measurement apparatus which projectively monitors its spin state in the $z$-direction. The ancilla $A$ can be driven by a control field along the $z$-axis. The target spins $B$ and $C$ interact exclusively with $A$. We assume that dephasing and relaxation effects from the environment do not affect the system, at least on the time scale of the protocol with optimal monitoring rate (see below). 
 The inter-spin coupling and the control field are adjusted by control functions $u_J(t)$ and $u_h(t)$, respectively, both assuming values in $[0,1]$.\\

\indent The Hamiltonian reads:
\begin{equation}\label{eq:sys_tot} 
H(t)=J_xu_J(t)\sigma_x^{\rm A}(\sigma_x^{\rm B}+\sigma_x^{\rm C})-h_z u_h(t)\sigma_z^{\rm A}\;,
\end{equation}
where $\sigma^i_j$ denote the Pauli spin operators ($j=x,y,z$) of spin $i$. 
Here the couplings $J_x$ and $h_z$ determine the magnitude of interaction and the field strength, respectively. 
The dynamics of the three qubit-system between consecutive measurements is induced by the Hamiltonian in Eq.~(\ref{eq:sys_tot}). Thus, the time evolution of the total density matrix $\rho$  of the tripartite system $A,B$, $C$ is governed by the Liouville-von Neumann equation  
\begin{eqnarray}\label{ME}
\dot{\rho}(t)=-\frac{i}{\hbar}[H(t),\rho(t)]\;.
\end{eqnarray}
Throughout the present work we scale energies and times with the coupling $J_x$ which can be in practice very small, favoring the isolation of the target qubits, but at the expense of a longer duration of the protocol. Accordingly, we consider for the control field driving the ancilla the (maximal) value $h_z=50~J_x$.\\
\indent The state of the three qubits (ancilla $A$ and target qubits $B,C$) are expressed in the basis $|nml\rangle:=|n_{\rm A}\rangle |m_{\rm B}\rangle|l_{\rm C}\rangle$, where $n,m,l\in\{0,1\}$ are the eigenvalues of the operators 
\begin{eqnarray}\label{zeta}
\hat{Z}^i=(\mathbf{1}^i+\sigma_z^i)/2
\;,
\end{eqnarray}
with $i=A,B$, and $C$, respectively.  We consider repeated \emph{projective} measurements of $\hat{Z}^{\rm A}$ which are assumed to be instantaneous, meaning that they take place on a time scale which is much smaller than the dynamical time scales of the system. Each measurement on the ancilla  projects the state of the full system into one of the eigensubspaces corresponding to the eigenvalues $0$ and $1$ of $\hat{Z}^{\rm A}$.\\  
\indent In the next sections we study the time evolution of the system subject to repeated  measurements of $\hat{Z}^{\rm A}$,  starting from  the fully separable initial state $\rho_0=|111\rangle\langle 111|$ with $h_z$ set to zero. After each measurement, the density matrix evolves unitarily with respect to $H(t)$, until the next measurement is performed.  The cycle is repeated for an overall time comprising several inter-measurement times.\\

\section{Evolution under nonselective monitoring and optimal inter-measurement time.}
\label{sec:nonselective}
\indent In our protocol, measurements of the spin state of the ancilla $A$ occur at equally-spaced times instants $t_n=n\tau$, where $\tau$ is the inter-measurement time and  $n\geq 1$. The scope of the present section is to establish how the time needed to eventually reach an asymptotic state   depends on $\tau$. For this purpose, we study the evolution of the system undergoing a sequence of \emph{nonselective} measurements in the absence of external fields, i.e., $u_h(t)=0$, and with $u_J(t)=1$.
In a nonselective measurement the outcome is disregarded or simply not available, thus yielding only a probabilistic information about the post-measurement state of the system. This is in distinct contrast to the case of \emph{selective} measurement  where each measurement prepares the system in the eigenstate corresponding to the outcome of the measured operator.
A nonselective measurement of $\hat{Z}^{\rm A}$ reduces the state of the entire system to a probabilistic mixture of projections  into the eigenstates $|0_{\rm A}\rangle$ and $|1_{\rm A}\rangle$ [see Eq.~(\ref{zeta})]. This transformation is provided by the action of the projectors $\Pi_i\equiv|i_{\rm A}\rangle\langle i_{\rm A}|\otimes\mathbf{1}_{\rm BC}$ with $i\in\{0,1\}$. Indeed, immediately after the $n$-th nonselective measurement, taking place at time $t_n=n\tau$, the  density matrix  of the total system can be written as 
\begin{eqnarray}\label{nonselective}
\rho(t_n)&=&\Pi_0U_n\rho(t_{n-1})U_n^\dag\Pi_0+\Pi_{1}U_n\rho(t_{n-1})U_n^\dag\Pi_{1}\;,
\end{eqnarray}
where $U_n$ is the time evolution operator from  $t_{n-1}$ to $t_n$.\\
\begin{figure}[htbp]
   \centering
   \includegraphics[width=3.2in]{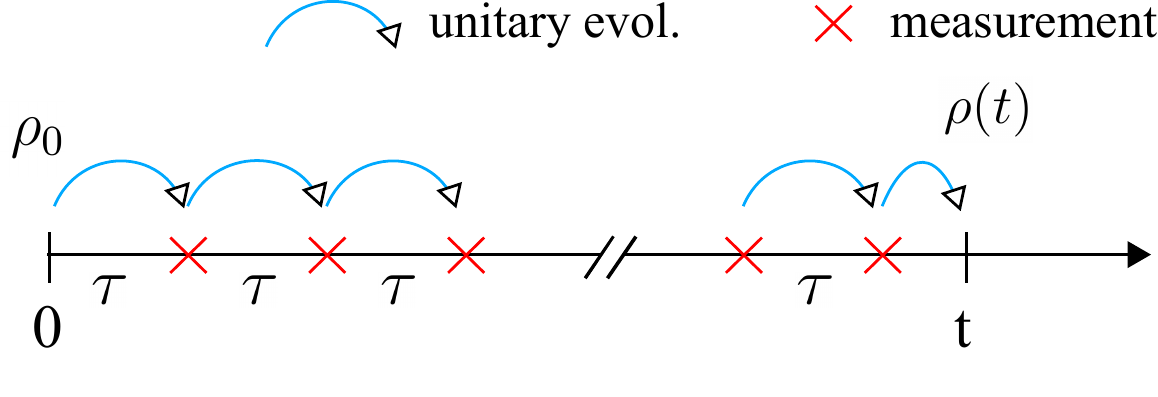} 
   \caption{Scheme of the unitary evolution of the full system interrupted by the sequence of equally spaced measurements of the state of qubit $A$. The time evolution between measurements is induced by the Hamiltonian~(\ref{eq:sys_tot}) according to the Liouville-von Neumann equation~(\ref{ME}).}  
   \label{fig:sequence}
\end{figure}

\begin{figure}[htbp]
   \centering
   \includegraphics[width=4in]{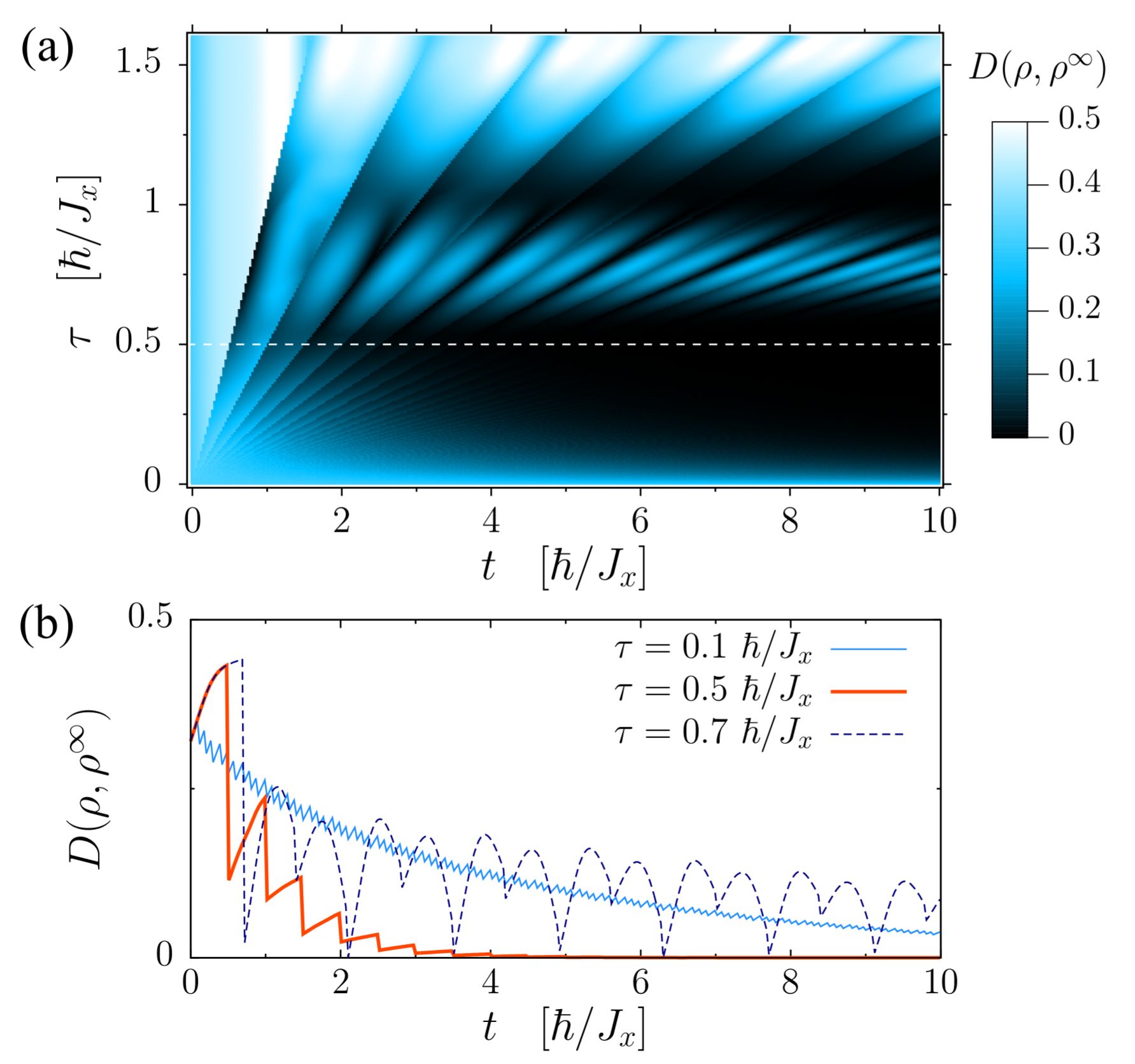}
   \caption{Time evolution of the full density matrix under repeated nonselective measurements. (a) Trace distance $D(\rho,\rho^\infty)$ between the time-evolved state $\rho\equiv\rho(t)$ and the asymptotic state $\rho^{\infty}$ [see Eq.~(\ref{eq:nsel})] as a function of the actual time $t$ and the inter-measurement time $\tau$. For each value of $\tau$ the density matrix is propagated up to the final time $t=10~\hbar/J_x$ according to the sequence depicted in Fig.~\ref{fig:sequence}. The time evolution of the density matrix between measurements is obtained by numerically solving of the Liouville-von Neumann equation~(\ref{ME}) with $u_h(t)=0$ and $u_J(t)=1$ in the Hamiltonian~(\ref{eq:sys_tot}). At the measurement times the state of the system is transformed as prescribed by Eq.~(\ref{nonselective}). The  initial state is $\rho(0)=|111\rangle\langle 111|$. The horizontal line at $\tau=0.5~\hbar/J_x$ highlights the optimal inter-measurement time to rapidly relax to the asymptotic state. In the limit  $\tau\rightarrow 0$ the system enters the quantum Zeno regime, as witnessed by the freezing of the trace distance from the asymptotic state at its initial value. (b) Trace distance {\it vs.} time for three fixed values of $\tau$. The red thick line corresponds to the optimal inter-measurement time.}
   \label{fig:taus}
\end{figure}

In the presence of repeated nonselective measurements, the density matrix at an arbitrary time $t$ is calculated as follows. Starting at the initial time $t_0=0$ in the state $\rho_0=|111\rangle\langle 111|$, the density matrix is propagated through Eq.~(\ref{ME}) for a time span $\tau$ after which the coherences between $A$ and the target qubits $B$ and $C$ are removed upon measuring the state of $A$, according to Eq.~(\ref{nonselective}). Then, the  post-measurement state $\rho(t_1)$ described by Eq.~(\ref{nonselective}) is used as the initial condition for a further propagation up to time $t_2=2\tau$ where a second measurement takes place. The above sequence is repeated and the total number of measurements occurring up to  time $t$ is given by the integer part of $t/\tau$. A scheme of  this sequence is depicted in Fig.~\ref{fig:sequence}.  We find that, for any non-pathological choice of the inter-measurement time $\tau$, i.e., for inter-measurement times that do not match multiples of the free system periodicity, the asymptotic state 
\begin{eqnarray}\label{eq:nsel}
\rho^{\infty}&=&\frac{1}{2}|1_{\rm A}\rangle\langle 1_{\rm A}|\otimes |\Phi^{-}_{\rm BC}\rangle\langle \Phi^{-}_{\rm BC}|+\frac{1}{4}|1_{\rm A}\rangle\langle 1_{\rm A}|\otimes|\Phi^{+}_{\rm BC}\rangle\langle \Phi^{+}_{\rm BC}|\nonumber\\
&&+\frac{1}{4}|0_{\rm A}\rangle\langle 0_{\rm A}|\otimes |\Psi^{+}_{\rm BC}\rangle\langle\Psi^{+}_{\rm BC}|
\end{eqnarray}
is eventually reached, where the four Bell states are defined as 
\begin{eqnarray}\label{eq:Bell}
|\Phi_{\rm BC}^{\pm}\rangle &=&\frac{1}{\sqrt{2}}\left(|11\rangle_{\rm BC}\pm|00\rangle_{\rm BC}\right)\nonumber\\
|\Psi_{\rm BC}^{\pm}\rangle &=&\frac{1}{\sqrt{2}}\left(|10\rangle_{\rm BC}\pm |01\rangle_{\rm BC}\right)\;.\quad
\end{eqnarray}
Note that the Bell state $|\Psi_{\rm BC}^-\rangle$ does not appear in Eq.~(\ref{eq:nsel}). This is due to the particular initial condition chosen, as explained in Appendix~\ref{App:nonsel}. Inspection of Eq.~(\ref{eq:nsel}) reveals that, once the asymptotic state is attained, a further measurement on $A$, with available outcome,  yields for the target qubits $B$ and $C$ the Bell state $|\Psi^{+}_{\rm BC}\rangle$ conditioned on the readout $0$, which occurs with probability $1/4$. On the other hand, the outcome $1$ yields for  the target qubits a probabilistic mixture of Bell states. 
In the following section and in Appendix~\ref{App:inspection},
a protocol that also yields the other entangled states $|\Psi^{-}_{\rm BC}\rangle$ and $|\Phi^{+}_{\rm BC}\rangle$ as pure states is specified.
\\
\indent To establish the optimal inter-measurement time $\tau^*$ for which the asymptotic state $\rho^\infty$ is reached in the least amount of time, we consider the trace-distance  between the time-evolved density operator $\rho(t)$ and the asymptotic state $\rho^\infty$. This quantity is defined as 
\begin{eqnarray}
D(\rho(t),\rho^\infty)&=&\frac{1}{2}{\rm Tr}|\rho(t)-\rho^{\infty}|\nonumber\\
&=&\frac{1}{2}\sum_i|\lambda_i|\;,
\end{eqnarray}
with $\lambda_i$ denoting the eigenvalues of the Hermitian matrix $\rho(t)-\rho^\infty$. \\
\indent The results for the trace  distance as a function of time $t$ and of the inter-measurement time $\tau$ are depicted in Fig.~\ref{fig:taus} for $u_h(t)=0$ (no external field on $A$) and $u_J(t)=1$ (constant qubits-ancilla coupling), see Eq.~(\ref{eq:sys_tot}). 
The plot in Fig.~\ref{fig:taus}(a) shows that an optimal inter-measurement time around the value $\tau^*=0.5~\hbar/J_x$ exists which corresponds to the minimal time $t$ needed to relax to the final state $\rho^\infty$ (see the horizontal dashed line). The curves depicted in Fig.~\ref{fig:taus}(b) clearly show that, for $\tau=\tau^*$, the asymptotic state is reached after few measurements whereas deviations from this optimal inter-measurement time entail a  considerably larger number of measurements. Indeed, in the limit of very small $\tau$ -- well below the optimal level -- the system enters the quantum Zeno regime, as witnessed by the freezing of the trace distance from the asymptotic state at its initial value (see the lower part of Fig.~\ref{fig:taus}(a))
\footnote{In a different setup~\cite{Nori2008},   measurements of the joint state of two qubits are proposed to entangle the qubits by exploiting the Zeno effect occurring at large measurement rates. Here, instead, we operate in the opposite regime where measurements at an appropriate, intermediate rate produce the fastest relaxation to a target state.}.
On the other hand, upon increasing the inter-measurement time $\tau$ above the optimal value $\tau^*$ the measurements eventually synchronize  with the periodicity of the free system so that oscillations persist for long evolution times around selected values of $\tau$, as shown in the upper part of Fig.~\ref{fig:taus}(a).

\section{Selective evolution under projective measurements and feedback}
\label{sec:selective}
Next we focus on the \emph{selective} evolution of the system under the same sequence of repeated measurements on $A$ detailed in Sec.~\ref{sec:nonselective} (see Fig.~\ref{fig:sequence}). In this case the individual realizations of the time evolution of $\rho(t)$ determined by specific sequences of random outcomes of the measurements on $A$ are considered. We show that the protocol converges to a factorized state with $|i_{\rm A}\rangle\langle i_{\rm A}|\otimes \rho^{_{\rm BC}}$, with the reduced system $BC$ in a Bell state.\\
\indent The evolution under selective measurements is calculated as follows: Starting at $t=0$ with the system in the state $\rho_0=|111\rangle\langle 111|$, the full density matrix is evolved according to Eq.~(\ref{ME}) for a time span $\tau$. At time $t_1=\tau$ the first measurement takes place yielding a random outcome $i\in\{0,1\}$ generated with probability ${\rm Tr}[\rho(\tau)\Pi_i]$. The state of the system is then updated to the post-measurement state which is in turn used as the initial condition for a further unitary evolution of duration $\tau$ and so on. The process, with the measurements taking place at  times $t_n=n\tau$, is depicted in Fig.~\ref{fig:sequence}. The explicit expression for the state immediately after the $n$-th measurement with outcome $i_n\in\{0,1\}$ is
\begin{equation}\label{eq:sel}
\rho(t_n)= \frac{\Pi_{i_n}U_n\rho(t_{n-1})U_n^\dag\Pi_{i_n}}{{\rm Tr}[U_n\rho(t_{n-1})U_n^\dag\Pi_{i_n}]}\;.
\end{equation}
The process described yields random realizations of the time evolution of the density matrix. The expression for the probability associated to a specific realization, i.e., to a specific readout sequence $(i_1,\dots,i_n,\dots)$, is given  in Appendix~\ref{App:inspection} [see Eq.~(\ref{eq:probseq})].
In the absence of a control field ($h_z=0$), we find two possible  behaviors:  Either the repeated measurements on $A$ yield an uninterrupted sequence of $1$'s, and the system is asymptotically left in the state $|1_{\rm 
A}\rangle|\Phi_{BC}^-\rangle$,  or the sequence randomly flips between the two outcomes $1$ and $0$. In the latter case  the state of the system flips between the two states $|1_{\rm A}\rangle|\Phi^+\rangle$ and $|0_{\rm A}\rangle|\Psi^+\rangle$. 
In Appendix~\ref{App:inspection} we account for this behavior by explicitly considering actual realizations of the selective evolution.\\
\begin{figure}[htbp] 
   \centering
    \includegraphics[width=6.2in]{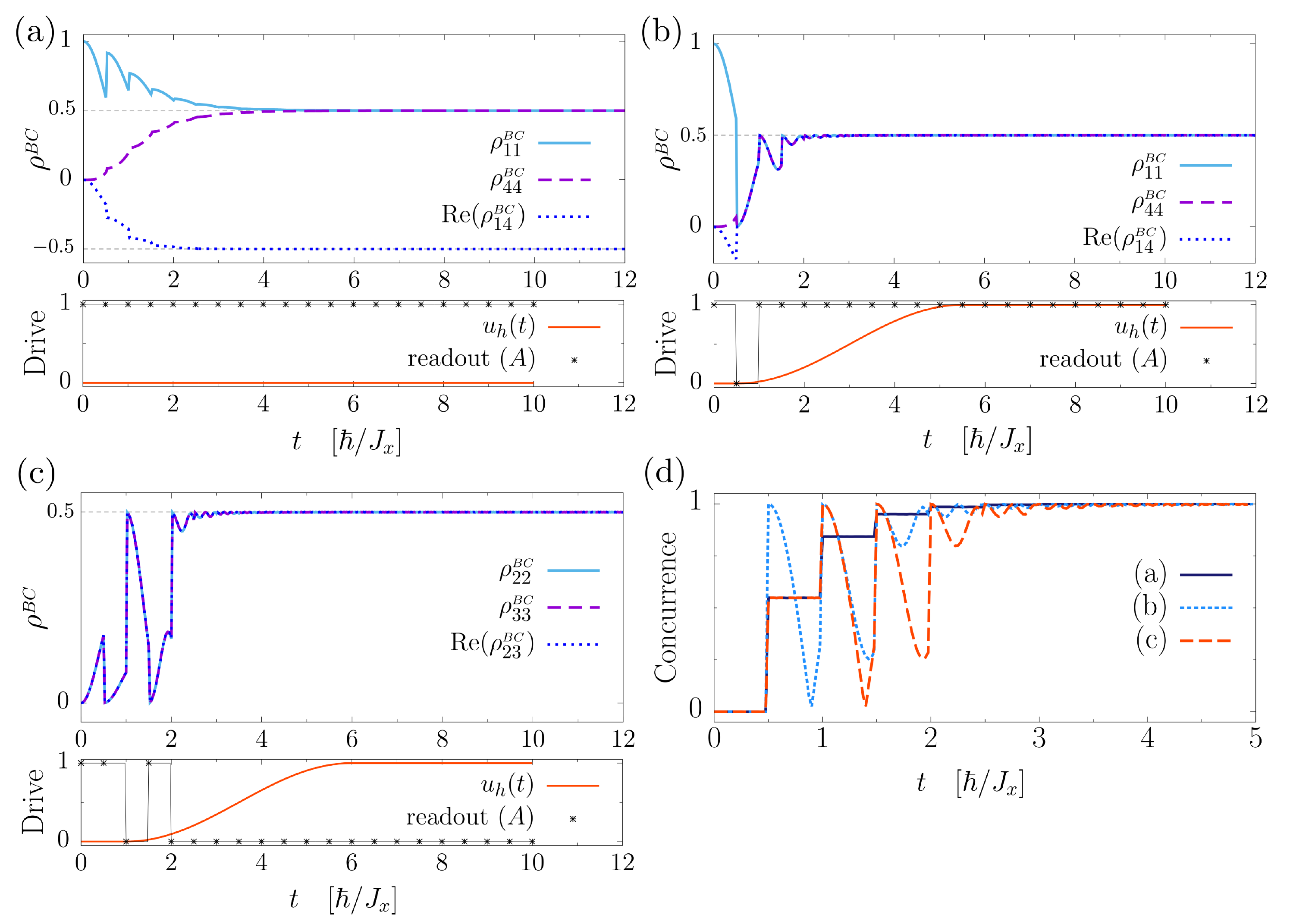} 
\caption{Time evolution of selected elements of the reduced density matrix $\rho^{_{\rm BC}}(t)$, for three sample realizations of the protocol with initial condition $|111\rangle$.  Each trajectory leads to a different Bell state for the target system $BC$. Specifically, (a) $|\Phi_{\rm BC}^{-}\rangle$, (b) $|\Phi_{\rm BC}^{+}\rangle$ and (c) $|\Psi_{\rm BC}^{+}\rangle$ (see Fig.~\ref{fig:scheme} below). Persistent fluctuations in the trajectories (b) and (c) are suppressed by applying an external control field of the form $-h_z u_h(t)$, with $h_z=50~J_x$, to the ancilla $A$. Below each evolution we depict the corresponding sequence of outcomes from measuring the state of $A$ and the time dependent function $u_h(t)$ of the control field [see~Eq.(\ref{eq:sys_tot})] acting on $A$. (d)  Time evolution of the concurrence of $\rho^{_{\rm BC}}(t)$ for the trajectories (a)-(c). The inter-measurement time is set to the optimal value $\tau=0.5~\hbar/J_x$.}
   \label{fig:traj}
\end{figure}
\indent To cope with the oscillating behavior taking place in a subset of the realizations of the protocol, a simple feedback scheme can be implemented which does not require particular  precision in its execution. A ramp of the control field $-h_z u_h(t)\sigma_z^{\rm A}$ acting on $A$ [see Eq.~(\ref{eq:sys_tot})] is turned on triggered by the first readout of a $0$ in the sequence of measurements.  Specifically, the field is initially zero and is then switched on at the time $t^*$ when a first detection of the state of $A$ with outcome $0$ occurs during the protocol. If no outcome $0$ is found the field stays off during the evolution [see for example panel (a) of Fig.~\ref{fig:traj} below]. Let $t_F$ be the total duration of the protocol. The switching function is the smooth ramp 
\begin{equation}\label{eq:ramp}
u_h(t)=
 \Bigg\{\begin{array}{lr}
\Theta(t-t^*)\{1-\cos[2\pi(t-t^*)/t_F]\}/2\;, \qquad &t\leq (t^*+t_F)/2 \\
1\;, \qquad &t> (t^*+t_F)/2 \\
\end{array}
\end{equation}
see Fig.~\ref{fig:traj}(a)-(c). 
 The stabilizing effect  on the oscillating sequences is due to the large value ($50~J_x$) of $h_z$  which makes the field term dominating with respect to the interaction term in the Hamiltonian~(\ref{eq:sys_tot}). As a result, a state with a definite eigenvalue of $\hat{Z}^{\rm A}$, such as a post-measurement state, is approximately an eigenstate of the Hamiltonian and does not evolve on the time scales of the protocol. We note that this stabilizing effect -- and thus the results in the present work -- does not depend on the precise form of the ramp $u_h(t)$. The final part of the protocol consists in switching off the interaction strength $J_x u_J(t)$: The interaction is held constant [$u_{J}(t)=1$] throughout the protocol and is switched off at time $t_F$, i.e., $u_{J}(t\geq t_F)=0$, after the target qubits $B$ and $C$ have reached a steady state.\\
\indent In Fig.~\ref{fig:traj}, the time evolution of the reduced density matrix $\rho^{_{\rm BC}}(t)$ under selective measurements  is shown for three sample realizations of the protocol. These realizations end up with the target qubits left in  different Bell states, corresponding to different sequences of the measurement readouts. The readout sequences are also shown for each sample time evolution along with the behavior of the control field acting on $A$. In the numerical simulations, the inter-measurement time $\tau$ is fixed at the optimal value $\tau^*=0.5~\hbar/J_x$ and the switch off takes place at $t_F=10~\hbar/J_x$, namely, after $20$ measurements. A realization of the protocol leaves the target qubits in the Bell state $|\Phi^{-}_{\rm BC}\rangle$ with probability $1/2$ or in one of the Bells states $|\Psi^{+}_{\rm BC}\rangle$ and $|\Phi^{+}_{\rm BC}\rangle$ with total probability $1/2$, (see the definitions in Eq.~(\ref{eq:Bell})).  Although the protocol generates the Bell states non-deterministically, once the steady state is reached they are unambiguously identified by reading the sequence of outcomes of the measurements on $A$. Having identified the Bell state encoded in the system $BC$, one can act \emph{locally} on qubit $B$ or $C$, by applying a rotation of the state of the qubit, to switch to a different Bells state~\cite{Preskill2004}. 
\indent A scheme of the protocol is provided in the upper panel of Fig.~\ref{fig:scheme}, where the three  realizations in Fig~(\ref{fig:traj})(a)-(c) are associated to the corresponding readout sequences and the final Bell state for the target qubits $B$ and $C$. Note that, by using a different initial condition,  the final Bell states associated to definite readout sequence can differ from the ones described here.  This is exemplified in Appendix~\ref{AppendixC}, where the protocol is carried out with the three qubits initially prepared in the state $|110\rangle$. 
\begin{figure}[htbp]
   \centering
   \includegraphics[width=3.5in]{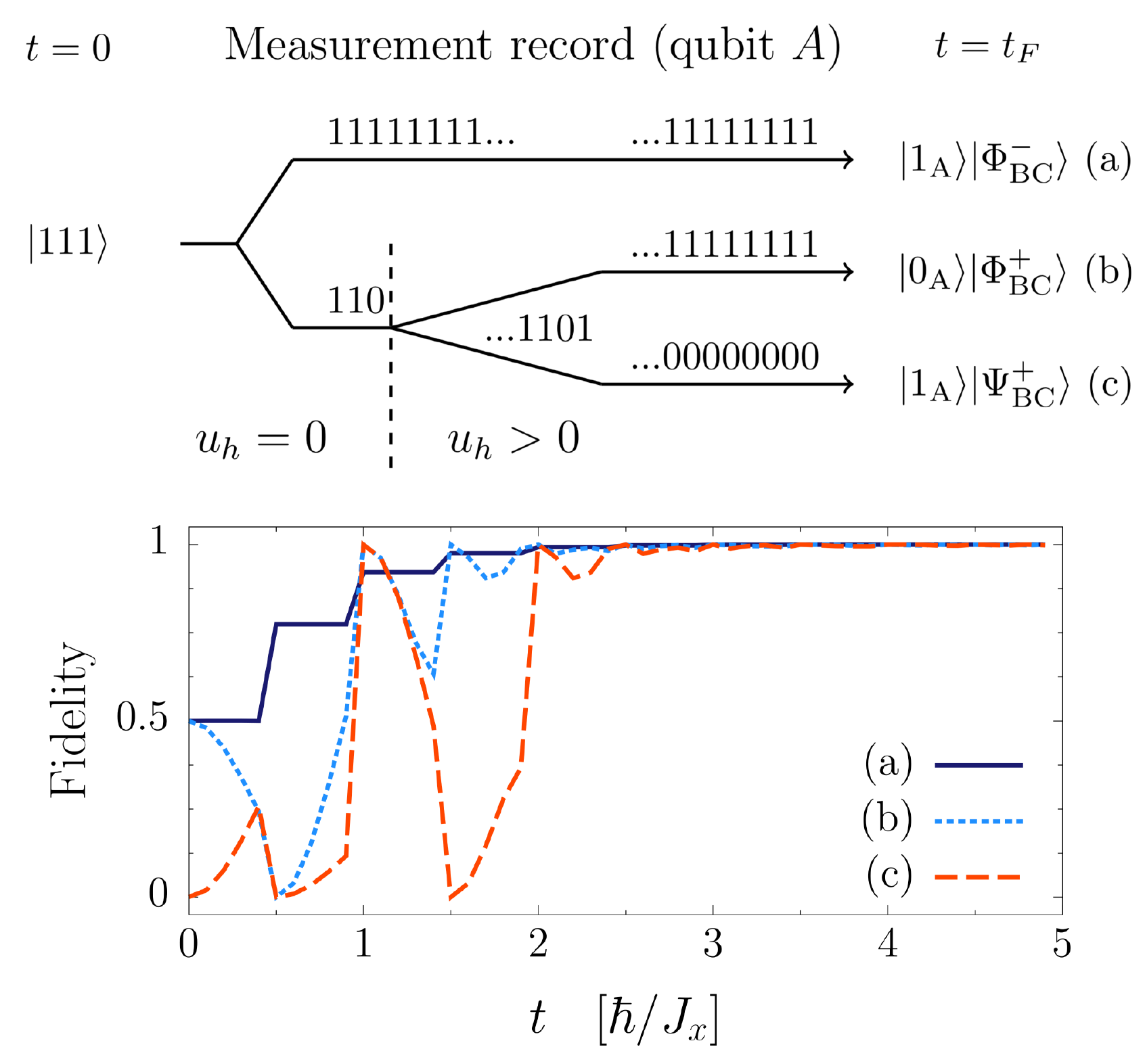} % [fig_2v1: width=4.5in] % [fig_2v2: width=5.5in]
   \caption{Upper panel -- Schematics of the protocol to generate Bell states in qubits $B$ and $C$ by a sequence of measurements on the ancilla qubit $A$, starting in the state $|111\rangle$. The scheme shows three different readout sequences with different resulting Bells states encoded in $BC$, corresponding to the three sample realizations in Fig.~\ref{fig:traj}(b)-(c). A smooth ramp of the control field $-h_zu_h(t)$, triggered by the first detection of a $0$ in the outcomes' sequence [see Fig.~\ref{fig:traj}], is applied to the ancilla in order to suppress the oscillatory behavior between $|1_A\rangle|\Phi_{\rm BC}^{+}\rangle$ and $|0_A\rangle|\Psi_{\rm BC}^{+}\rangle$ and converge to a definite result. Lower panel -- Fidelity of the trajectories depicted in Fig.~\ref{fig:traj}(a)-(c) to the corresponding target Bell states.}
   \label{fig:scheme}
\end{figure}

To study the degree of entanglement during the protocol, we evaluate the concurrence of the reduced density matrix of the target system $BC$. This quantity is defined as $\mathcal{C}[\rho^{_{\rm BC}}(t)]= {\rm max}(0,\lambda_1-\lambda_2-\lambda_3-\lambda_4)$, where the $\lambda_i$'s are the ordered eigenvalues of the matrix $\rho^{_{\rm BC}}(t)(\sigma_y^{\rm B}\otimes\sigma_y^{\rm C})[\rho^{_{\rm BC}}(t)]^\ast(\sigma_y^{\rm B}\otimes\sigma_y^{\rm C})$ \cite{Wootters97,Wootters98}. The concurrence takes values between zero and one, corresponding to non-entangled (zero) and maximally entangled states (one), respectively. The results are depicted in Fig.~\ref{fig:traj}(d) for the three sample realizations displayed in panels (a)-(c). 
%The latter expression is valid for both pure and mixed states and hence for selective and non-selective measurement sequences. 
In addition, we confirm  that the reduced two-qubit system $BC$ is left in the specific Bell state corresponding to a particular readout sequence. This is done by calculating the fidelity~\cite{Uhlmann76,Jozsa94}, given by the expression
\begin{equation}
F[\rho^{_{\rm BC}}(t),\rho^{_{\rm BC}}_\infty] = {\rm Tr}\left[\sqrt{\sqrt{\rho^{_{\rm BC}}(t)}\rho^{_{\rm BC}}_\infty\sqrt{\rho^{_{\rm BC}}(t)}}\right]\;,
\end{equation}
where the referential density matrix $\rho^{_{\rm BC}}_\infty$ is chosen \emph{a posteriori}, as the final state that corresponds to the readout sequence (see the upper panel of Fig.~\ref{fig:scheme}). The time evolutions of the fidelity for the sample evolutions in Fig.~\ref{fig:traj} are depicted in the lower panel of  Fig.~\ref{fig:scheme}.\\
\indent We conclude this section by noting that, in a realistic situation,  environmental effects are present, especially on the ancilla qubit $A$ which is connected to the meter. This  makes the choice of the optimal time crucial in order to conclude the protocol before these detrimental effects spoil it. However,  the realizations of the protocol in the classes depicted in Fig.~\ref{fig:traj}(a) and (b) are expected to be more robust with respect to the environmental influence, as compared to those where the large control field acting on $A$ freezes the system  in the higher energy state $|0_{\rm A}\rangle|\Psi_{\rm BC}^+\rangle$. In this latter case, the large energy splitting may cause the decay $|0_{\rm A}\rangle\rightarrow|1_{\rm A}\rangle$ before the protocol is completed. \\
\indent Alternatively, one may think of freezing the oscillations between $|1_{\rm A}\rangle|\Phi^+\rangle$ and $|0_{\rm A}\rangle|\Psi^+\rangle$ by the Zeno effect, namely by suddenly let the inter-measurement time $\tau$ 
go to zero after a measurement on the ancilla with outcome $0$.
\section{Conclusions}

With this work a simple protocol is presented for generating Bell states in a couple of  qubits which do not interact directly. The qubits are entangled by means of  repeated projective measurements of the state of a shared ancilla qubit. We have shown that the protocol yields a definite and stable Bell state, starting from  the completely factorized state $|111\rangle$ of the full system. This is attained by acting on the ancilla
with a suitable form of feedback, namely a ramp of a control field [see Eq.(\ref{eq:ramp})], triggered by a specific readout of the ancilla state. The backaction of the  repeated measurements asymptotically yields a stationary state of the form $|i_{\rm A}\rangle|\varphi_{\rm BC}\rangle$, with $i \in \{0,1\}$ and $|\varphi_{\rm BC}\rangle\in\{|\Phi_{\rm BC}^{-}\rangle,\;|\Psi_{\rm BC}^{+}\rangle,\; |\Phi_{\rm BC}^{+}\rangle \}$, where the Bell state is unambiguously identified by reading the sequence of measurement outcomes of the ancilla.
 
\section*{Acknowledgments}
This work is dedicated to Wolfgang~P.~Schleich, a great scientist and constant source of inspiration, on the occasion of his 60$^{\rm th}$ birthday. P.~H. and P.~T. also thank Wolfgang for his vivid and illuminating many discussions and constructive debates during our yearly, alternating Augsburg-Ulm workshops.  In addition P.~H. likes to thank Kathy and Wolfgang for their lifetime-long wonderful friendship.

\appendix
%\vspace{2cm}
\section{Asymptotic state for  nonselective repeated measurements with different initial conditions}
\label{App:nonsel}
To gain insight in the reasons why, with the chosen initial separable state 
\begin{equation}
|111\rangle=\frac{1}{\sqrt{2}}|1_{\rm A}\rangle(|\Phi_{\rm BC}^+\rangle+|\Phi_{\rm BC}^-\rangle)\;,
\end{equation}
the sequence of nonselective measurement considered in Sec.~\ref{sec:nonselective} ends up with asymptotic  state in Eq.~(\ref{eq:nsel}), it is sufficient  
to note that the  Hamiltonian~(\ref{eq:sys_tot}), in the absence of control fields and with constant interaction strength,  can be written as
\begin{equation}\label{eq:H_Bell}
H(t)=2J_x\sigma_x^{\rm A}(|\Phi^{+}_{\rm BC}\rangle\langle\Psi^{+}_{\rm BC}|+|\Psi^{+}_{\rm BC}\rangle\langle\Phi^{+}_{\rm BC}|)\;.
\end{equation}
The state $|1_{\rm A}\rangle|\Phi_{\rm BC}^-\rangle$ is an eigenstate of the Hamiltonian and the only states involved that are connected by a nonzero transition amplitude are $|1_{\rm A}\rangle|\Phi_{\rm BC}^+\rangle$ and $|0_{\rm A}\rangle|\Psi_{\rm BC}^+\rangle$. Thus, with the chosen initial state,  the probability to get the target qubits in the Bell state $|\Psi_{\rm BC}^-\rangle$ is zero, as the dynamics induced by Hamiltonian~(\ref{eq:H_Bell}) is confined to the subspace spanned by 
$\{|1_{\rm A}\rangle|\Phi_{\rm BC}^-\rangle,\;|1_{\rm A}\rangle|\Phi_{\rm BC}^+\rangle,\;|0_{\rm A}\rangle|\Psi_{\rm BC}^+\rangle\}$ and the measurements on $A$ do not affect this feature.
 On the other hand, by starting with a different   initial state, as for example $|110\rangle=|1_{\rm A}\rangle(|\Psi_{\rm BC}^+\rangle+|\Psi_{\rm BC}^-\rangle)/\sqrt{2}$, the asymptotic state reads
\begin{eqnarray}\label{rho_as_Appendix}
\rho^{\infty}&=&\frac{1}{2}|1_{\rm A}\rangle\langle 1_{\rm A}|\otimes |\Psi^{-}_{\rm BC}\rangle\langle \Psi^{-}_{\rm BC}|+\frac{1}{4}|1_{\rm A}\rangle\langle 1_{\rm A}|\otimes|\Psi^{+}_{\rm BC}\rangle\langle \Psi^{+}_{\rm BC}|\nonumber\\
&&+\frac{1}{4}|0_{\rm A}\rangle\langle 0_{\rm A}|\otimes |\Phi^{+}_{\rm BC}\rangle\langle\Phi^{+}_{\rm BC}|\;,
\end{eqnarray}
[cf Eq.~(\ref{eq:nsel})]. A complete list of the asymptotic states attained by starting in each of the (computational) basis states is shown in Table~\ref{tab:cases}.  \\
\begin{table}[ht!]
  \centering
\small{
\begin{tabular}{ |l|l| }
  \hline
init. state & \hspace{2cm} $\rho^{\infty}$\\
  \hline
$\begin{array}{lr}
|111\rangle \\
|100\rangle \\
\end{array}$
& Eq.~(\ref{eq:nsel})\\
  \hline
$\begin{array}{lr}
|110\rangle \\
|101\rangle \\
\end{array}$
& Eq.~(\ref{rho_as_Appendix})\\
  \hline
$\begin{array}{lr}
|011\rangle \\
|000\rangle \\
\end{array}$
& Eq.~(\ref{eq:nsel}) with $|1_{\rm A}\rangle\leftrightarrow |0_{\rm A}\rangle$\\
  \hline
$\begin{array}{lr}
|010\rangle \\
|001\rangle \\
\end{array}$
& Eq.~(\ref{rho_as_Appendix}) with $|1_{\rm A}\rangle\leftrightarrow |0_{\rm A}\rangle$\\
\hline
\end{tabular}
}
\caption{Repeated nonselective measurements in the absence of external fields -- asymptotic mixed density matrix for eight initial configurations.}
\label{tab:cases}
\end{table}
To gain an intuition on how the entries in the table are obtained, let us introduce the map $\mathcal{M}$ that propagates the density matrix for a time span $\tau$ and then applies a nonselective measurement on the ancilla. This map is defined by the action
\begin{eqnarray}\label{nselApp}
\rho(t_n)&=&\Pi_0U_n\rho(t_{n-1})U_n^\dag\Pi_0+\Pi_{1}U_n\rho(t_{n-1})U_n^\dag\Pi_{1}\nonumber\\
&\equiv& \mathcal{M}[\rho(t_{n-1})],
\end{eqnarray}
Then, the asymptotic state is given by 
\begin{eqnarray}\label{rhoinfApp}
\rho^\infty&=&\lim_{n\rightarrow\infty}\mathcal{M}^n[\rho_0]\;.
\end{eqnarray}
Now, let $\mathcal{F}_j$ be the operation that flips the state of spin $j\in\{A,B,C\}$ in the computational basis (the eigenbasis of $\hat{Z}^j$). For $j\in\{B,C\}$, both the Hamiltonian and the projectors $\Pi_i$ are invariant under such operation, we have
\begin{eqnarray}\label{M}
\mathcal{M}^n[\mathcal{F}_{\rm B}\rho_0]=\mathcal{F}_{\rm B}\mathcal{M}^n[\rho_0]\;,
\end{eqnarray}
and similarly $\mathcal{M}^n[\mathcal{F}_{\rm C}\rho_0]=\mathcal{F}_{\rm C}\mathcal{M}^n[\rho_0]$ and $\mathcal{M}^n[\mathcal{F}_{\rm B}\mathcal{F}_{\rm C}\rho_0]=\mathcal{F}_{\rm B}\mathcal{F}_{\rm C}\mathcal{M}^n[\rho_0]$. It follows that flipping   the spin of either $B$ or $C$ in the initial state $\rho_0=|111\rangle\langle 111|$ returns the asymptotic state in Eq.~(\ref{eq:nsel}) with the spin of $B$ or $C$ flipped, i.e., Eq.~(\ref{rho_as_Appendix}). On the other hand, flipping both the spins of $B$ and $C$ in the initial state returns the asymptotic state~(\ref{eq:nsel}) itself, because $\mathcal{F}_{\rm B}\mathcal{F}_{\rm C}\rho^{\infty}=\rho^{\infty}$. The above reasoning accounts for the first two rows of Table~\ref{tab:cases}. \\
\indent In a similar way, by writing $\sigma^{\rm A}=|0_{\rm A}\rangle\langle 1_{\rm A}|+|1_{\rm A}\rangle\langle 0_{\rm A}|$ we see that the Hamiltonian is invariant under the action of $\mathcal{F}_{\rm A}$ and that $\mathcal{F}_{\rm A}\Pi_{0(1)}=\Pi_{1(0)}$, which entails  $\mathcal{M}^n[\mathcal{F}_{\rm A}\rho_0]=\mathcal{F}_{\rm A}\mathcal{M}^n[\rho_0]$ [see Eq.~(\ref{nselApp})]. This accounts for the last two rows of Table~\ref{tab:cases}.

%\clearpage

\section{Details of the selective evolution}\label{App:inspection}

Let us inspect the actual realizations of the time evolution given by repeatedly measuring the state of $A$ in a selective fashion, starting from the   state $|111\rangle$ in the absence of control fields, $u_h(t)=0$. The action of the time evolution operator induced by the Hamiltonian~(\ref{eq:H_Bell})  for a time span $\tau$ is 
\begin{equation}\label{eq:U}
U(\tau)|111\rangle=\frac{1}{\sqrt{2}}|1_{\rm A}\rangle\left(|\Phi_{\rm BC}^-\rangle+a|\Phi_{\rm BC}^+\rangle\right)+\frac{b}{\sqrt{2}}|0_{\rm A}\rangle|\Psi_{\rm BC}^+\rangle\;,
\end{equation}
with $|a|^2+|b|^2=1$, where $a$ and $b$ depend on $\tau$.
The first measurement on the ancilla collapses this evolved state into one of the two alternative outcomes
\begin{equation}
\begin{array}{lr}
|\phi_{(1)}\rangle =|1_{\rm A}\rangle(|\Phi_{\rm BC}^-\rangle+a|\Phi_{\rm BC}^+\rangle)/\mathcal{N}_1\qquad{\rm prob}=\mathcal{N}_1^2/2&\\
|\phi_{(0)}\rangle =|0_{\rm A}\rangle|\Psi_{\rm BC}^+\rangle\;\qquad\qquad\qquad\qquad{\rm prob}=|b|^2/2 &\\
\end{array}\;,
\end{equation}
where the indexes $(1/0\dots)$ are used for bookkeeping the sequence of outcomes and $\mathcal{N}_n=\sqrt{1+|a|^{2n}}$. Then, evolving the above states for another time span $\tau$ we get
\begin{eqnarray}
U(\tau)|\phi_{(1)}\rangle &=&|1_{\rm A}\rangle\left(|\Phi_{\rm BC}^-\rangle+a^2|\Phi_{\rm BC}^+\rangle\right)/\mathcal{N}_1+ab|0_{\rm A}\rangle|\Psi_{\rm BC}^+\rangle/\mathcal{N}_1\\
U(\tau)|\phi_{(0)}\rangle &=& a|0_{\rm A}\rangle|\Psi_{\rm BC}^+\rangle+b|1_{\rm A}\rangle|\Phi_{\rm BC}^+\rangle
\end{eqnarray}
Upon measuring again the state of $A$
the following two couples of alternative outcomes arise
\begin{eqnarray}
\Bigg\{\begin{array}{lr}
|\phi_{(11)}\rangle =|1_{\rm A}\rangle(|\Phi_{\rm BC}^-\rangle+a^2|\Phi_{\rm BC}^+\rangle)/\mathcal{N}_2\qquad{\rm prob}=\mathcal{N}_2^2/\mathcal{N}_1^2&\\
|\phi_{(10)}\rangle=|0_{\rm A}\rangle|\Psi_{\rm BC}^+\rangle\;\qquad\qquad\qquad\qquad\;{\rm prob}=|ab|^2/\mathcal{N}_1^2&\\
\end{array}\\
\Bigg\{\begin{array}{lr}
|\phi_{(01)}\rangle =|1_{\rm A}\rangle|\Phi_{\rm BC}^+\rangle\quad\qquad\qquad\qquad\qquad{\rm prob}=|b|^2&\\
|\phi_{(00)}\rangle=|0_{\rm A}\rangle|\Psi_{\rm BC}^+\rangle\qquad\qquad\qquad\qquad\quad{\rm prob}=|a|^2&\\
\end{array}\label{eq:b6}\;,
\end{eqnarray}
and so on. From this behavior it is clear that, for $a\neq 1$, i.e., if $\tau$ is not a multiple of the period of the unitary evolution, a long sequence of outcomes $1$ will yield the stabilized state $|1_{\rm A}\rangle|\Phi_{\rm BC}^-\rangle$, meaning that a further measurement  will leave the system in this same  state with probability $\sim 1$. On the other hand, the presence of zeroes in the readout sequences entails oscillations between $|1_{\rm A}\rangle|\Phi_{\rm BC}^+\rangle$ and $|0_{\rm A}\rangle|\Psi_{\rm BC}^+\rangle$ which last indefinitely. This behavior is very close to what is found in Ref~\cite{Buttiker06} by simulating the repeated parity measurements in a couple of double quantum dot qubits.  The probability associated to a specific sequence of outcomes $(i_1,i_2,...,i_N)$, where $i_n\in\{0,1\}$, is given by
\begin{equation}
P(i_1,i_2,...,i_N)={\rm Tr}[\Pi_{i_{N}}U_{N}...\Pi_{i_2}U_2\Pi_{i_1}U_1\rho_0U^{\dag}_1\Pi_{i_1}U^{\dag}_2\Pi_{i_2}...U^{\dag}_{N}\Pi_{i_{N}}]\;,
\end{equation}
where, as in Eq.~(\ref{nonselective}), $U_n$ is the time evolution operator from  $t_{n-1}$ to $t_n$. By extrapolating from the sequence in Eqs.~(\ref{eq:U})-(\ref{eq:b6}) one sees that the probability to obtain an interrupted sequence of $n$ readouts $1$ is\\
\begin{equation}\label{eq:probseq}
P(1,\dots,1)=\frac{\mathcal{N}_1}{2}\frac{\mathcal{N}_2}{\mathcal{N}_1}\dots\frac{\mathcal{N}_n}{\mathcal{N}_{n-1}}=\frac{1}{2}\frac{\mathcal{N}_n}{\mathcal{N}_{n-1}}\quad\stackrel{n\gg 1}{\longrightarrow}\quad \frac{1}{2}\;.
\end{equation}

%\clearpage
\section{Realizations of the protocol with a different initial condition}\label{AppendixC}
\begin{figure}[htbp] 
   \centering
    \includegraphics[width=6.5in]{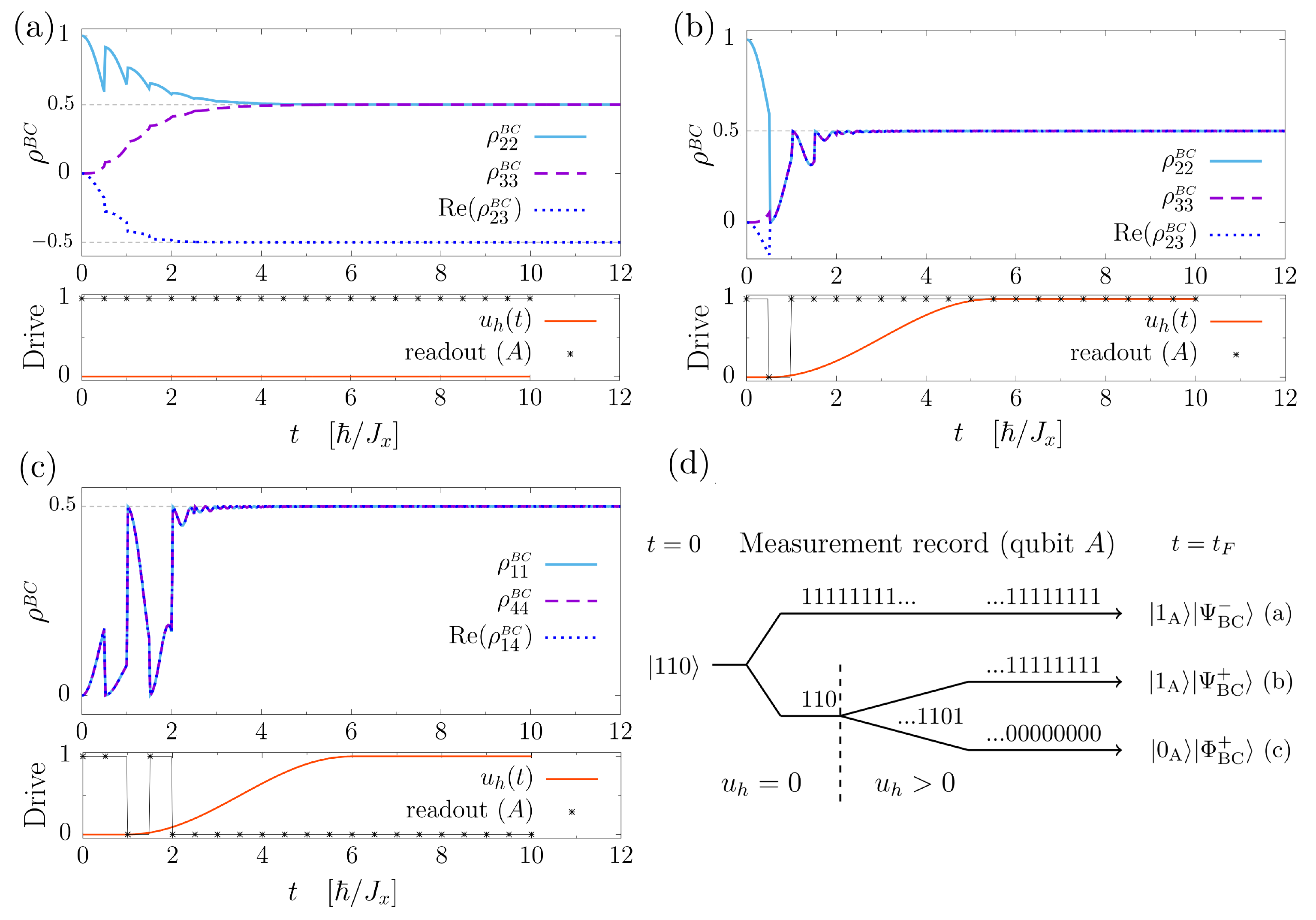} 
\caption{Time evolution of selected elements of the reduced density matrix $\rho^{_{\rm BC}}(t)$ and scheme of the protocol. Each of the three sample realizations of the protocol starts with the system in the state $|110\rangle$ and leads to a different Bell state for the target system $B,C$. Specifically, (a) $|\Psi_{\rm BC}^{-}\rangle$, (b) $|\Psi_{\rm BC}^{+}\rangle$ and (c) $|\Phi_{\rm BC}^{+}\rangle$. Persistent fluctuations in the trajectories (b) and (c) are suppressed by applying an external control field of the form $-h_z u_h(t)$, with $h_z=50~J_x$, to the ancilla $A$. Below each evolution we depict the corresponding sequence of outcomes from measuring the state of $A$ and the time dependent control function $u_h(t)$. The inter-measurement time is set to the optimal value $\tau=0.5~\hbar/J_x$. (d) Scheme of the readout sequences and final Bell states corresponding to panels (a)-(c) (cf Fig.~\ref{fig:scheme}).}
   \label{fig:trajAppendix}
\end{figure}
\indent In Fig.~\ref{fig:trajAppendix} we depict three sample realizations of the protocol detailed in Sec.~\ref{sec:selective} starting in the factorized state $|110\rangle=|1_{\rm A}\rangle(|\Psi_{\rm BC}^+\rangle+|\Psi_{\rm BC}^-\rangle)/\sqrt{2}$. A schematics of the protocol's outcomes and readout sequences is also shown. 

\clearpage

\bibliographystyle{unsrt} %plain, unsrt
%\bibliography{biblio}

\end{document}